\documentclass[a4paper]{article}

\usepackage{ISCSLP_v2}
\usepackage{mathrsfs}
\usepackage{verbatim}
\usepackage{multirow}
\usepackage{siunitx}

\title{End-to-end Language Identification using NetFV and NetVLAD}

\name{Jinkun Chen$^2$, Weicheng Cai$^{1,2}$, Danwei Cai$^1$, Zexin Cai$^1$, Haibin Zhong$^3$, Ming Li$^1$\thanks{This research was funded in part by the National Natural Science Foundation of China (61773413), Natural Science Foundation of Guangzhou City (201707010363), National Key Research and Development Program (2016YFC0103905).}}

\address{
	$^1$Data Science Research Center, Duke Kunshan University, Kunshan, China \\
	$^2$School of Electronics and Information Technology, Sun Yat-sen University, Guangzhou, China
	$^3$Jiangsu Jinling Science and Technology Group Limited, Nanjing, China
}
\email{ming.li369@dukekunshan.edu.cn}

\begin{document}
	
	\maketitle
	\def\bx{\boldsymbol{x}}
	\def\bX{\boldsymbol{X}}
	\def\bsigma{\boldsymbol{\sigma}}
	\def\bmu{\boldsymbol{\mu}}

	\begin{abstract}
		In this paper, we apply the NetFV and NetVLAD layers  for the end-to-end language identification task. NetFV and NetVLAD layers are the differentiable implementations of the standard Fisher Vector
		and Vector of Locally Aggregated Descriptors (VLAD) methods, respectively. Both of them can encode  a sequence of feature vectors into a fixed dimensional vector which is very important to process those variable-length utterances. 
		We first present the relevances and differences between the classical i-vector and the aforementioned encoding schemes. 
		Then, we construct a flexible end-to-end framework including a convolutional neural network (CNN) architecture and an encoding layer (NetFV or NetVLAD) for the language identification task.
		Experimental results on the NIST LRE 2007 close-set task show that the proposed system achieves significant EER reductions against the conventional i-vector baseline and the CNN temporal average pooling system, respectively. 
	\end{abstract}
	
	\section{Introduction}
	Language identification (LID) is a kind of utterance-level paralinguistic speech attribute recognition task with variable-length sequences as inputs.
	For the input utterances, the duration might range from a few seconds to several minutes. Besides, there are no constraints on the lexical words thus the training utterances and test segments may have phonetic mismatch issue~\cite{kinnunen2010overview}. Therefore, our purpose is to find an effective and robust method to retrieve the utterance-level information and encode them into fixed dimensional vector representations.
	
	To address the variable-length inputs issue for acoustic feature based LID, many methods have been proposed in the last two decades.
	The deterministic Vector Quantization (VQ) model is used for LID in~\cite{cimarusti1982development, sugiyama1991automatic}.
	VQ assigns the frame-level acoustic features to the nearest cluster in codebook and calculates the VQ distortions. Every language is characterized by an occupancy probability histogram.
	Compared to VQ, the Gaussian Mixture Model (GMM) 
	is capable to model the complex distribution of the acoustic features~\cite{zissman1996comparison} and generates soft posterior probabilities to assign those frame-level features to Gaussian components.
	Once the GMM is trained, the zero-order and first-order Baum-Welch statistics can be accumulated to construct a high dimensional GMM Supervector~\cite{kinnunen2010overview,campbell2006support, you2010gmm}, which is considered as an utterance-level representation. 
	Furthermore, the GMM Supervector can be projected to a low rank subspace  using the factor analysis technique, which results in the i-vector~\cite{dehak2011front, dehak2011language}. 
	Recently, the posterior statistics on each decision tree senones generated by the speech recognition acoustic model are adopted to construct the phonetic aware DNN i-vector~\cite{yunlei2014DNNivector}, which outperforms the GMM i-vector in LID tasks~\cite{li2014speaker, snyder2015time, richardson2015deep}  due to the discriminative frame level alignments.
	
	More recently, some end-to-end learning approaches are proposed for the LID task and achieved superior performances~\cite{lopez2014automatic, gonzalez2014automatic}. However, these methods may lack the flexibility in dealing with duration-variant utterances. This is mainly because the deep learning module (fully-connected layer) or the development platform usually requires fixed-length inputs. 
	Each utterance with different number of frames has  to be zero padded  or truncated into fixed-size vectors. This is not a desired way to recognize spoken languages, speaker identities or other paralinguistic attributes on speech utterances with various durations. 
	To address this problem, recurrent neural networks (RNN), e.g. Long Short-Term Memory (LSTM)~\cite{hochreiter1997long}, is introduced to LID and the last time-step output of the RNN layer is used as the utterance-level representation~\cite{gelly2016divide}. 
	Stacked long-term time delay neural networks (TDNN) is adopted to span a wider temporal context on the inputs and a hierarchical structure is built to predict the likelihood over different languages~\cite{garcia2016stacked}.
	Alternatively, modules like temporal average pooling (TAP), are proposed to perform statistic measures on the length-variant feature sequences in order to generate fixed dimensional representations for LID~\cite{snyder2016deep, li2017deep}.
	These end-to-end systems perform well, however, the simple statistic measures are performed globally on all the frame level features (e.g. average pooling) which may smooth out the information on each clusters.
	In our early works, we imitate the GMM Supervector encoding procedure and introduce a learnable dictionary encoding (LDE) 
	layer for the end-to-end LID   system~\cite{caiwch2018lde,caiwch2018e2elre}. 
	The success of LDE layer in the end-to-end LID framework inspires us to explore different encoding methods which may be feasible for the LID task.
	
	In this paper, we adopt NetFV~\cite{tang2016deep} and NetVLAD~\cite{arandjelovic2016netvlad} in our end-to-end LID task and explore the feasibility of this two methods.
	NetFV is the ``soft assignment" version of standard Fisher Vector (FV)~\cite{jaakkola1999exploiting, perronnin2007fisher, perronnin2010large} and is differentiable which could be easily integrated to an end-to-end trainable system. Meanwhile, Vector of Locally Aggregated Descriptors (VLAD) proposed in ~\cite{jegou2010aggregating} is a simplified non-probabilistic version of the standard FV and similarly, 
	VLAD is further enhanced as a trainable layer named NetVLAD in ~\cite{arandjelovic2016netvlad}. Standard FV and VLAD have been widely employed in computer vision tasks such as image retrieval, place recognition and video classification~\cite{perronnin2010large, jegou2010aggregating, sanchez2013image}.
	Moreover, both NetFV and NetVLAD are considered as more powerful pooling techniques to aggregate variable-length inputs into a fixed-length representation. This two encoding layers have been widely used and perform well in vision tasks~\cite{tang2016deep, arandjelovic2016netvlad, miech2017learnable}.
	As for the LID task, we employ a residual networks (ResNet)~\cite{he2016deep} as the front-end feature extractor, and use NetFV or NetVLAD to encode the variable-size CNNs feature maps into fixed-size utterance-level representations. Experimental results on NIST LRE 07 show that the proposed method outperforms the  GMM i-vector baseline as well as the TAP layer based end-to-end systems.
	Moreover, the proposed end-to-end system is flexible, effective and robust in both training and test phases.
	
	The following of this paper is organized as follows: Section~\ref{sec:methods} explains the LID methods based on the GMM i-vector, NetFV and NetVLAD  as well as  the overall end-to-end framework.
	Experimental results and discussions are presented in Section~\ref{sec:experiments} while conclusions and future works are provided in Section~\ref{sec:conclusions}.
	
	\section{Methods}
	\label{sec:methods}
	In this section, we elaborate the mechanisms of the GMM Supervector, NetFV and NetVLAD. Besides, we explain the relevances and differences of these three encoding schemes. Furthermore, we describe our flexible end-to-end framework in details.
	
	\subsection{GMM Supervector}
	Given a $C$-component Gaussian Mixture Model-Universal Background Model (GMM-UBM) with parameters set $\lambda=\{\alpha_c, \boldsymbol{\mu}_c, \boldsymbol{\Sigma}_c, c=1,2,\dots,C\}$, where $\alpha_c$, $\boldsymbol{\mu}_c$ and  $\boldsymbol{\Sigma}_c$ are the  mixture  weight, mean vector and covariance matrix of the $c^{th}$ Gaussian component, respectively, and a $L$-frame speech utterance with $D$ dimensional features $\boldsymbol{X}=\{\boldsymbol{x}_i, i=1,2,\dots,L\}$, 
	the normalized $c^{th}$ component's Supervector is defined as	
	\begin{equation}
	\label{eq:supervector}
	\widetilde{\boldsymbol{F}}_c=\frac{\boldsymbol{F}_c}{N_c}=\frac{\sum_{t=1}^L{P(c|\boldsymbol{x}_t, \lambda)}\cdot(\boldsymbol{x}_t-\boldsymbol{\mu}_c)}{\sum_{l=1}^L{P(c|\boldsymbol{x}_l, \lambda)}},
	\end{equation}
	where $P(c|\bx_c,\lambda)$ is the occupancy probability for $\boldsymbol{x}_t$ on the $c^{th}$ component of the GMM, the numerator $\boldsymbol{F}_c$ and the denominator $N_c$ are referred as the zero-order and first-order centered Baum-Welch statistics.
	By concatenating all of the $\widetilde{\boldsymbol{F}}_c$ together, we derive the high dimensional Supervector $\widetilde{\boldsymbol{F}} \in \mathbb{R}^{CD\times1}$ of the corresponding utterance is
	\begin{equation}
	\widetilde{\boldsymbol{F}} = [\widetilde{\boldsymbol{F}}_1^T, \widetilde{\boldsymbol{F}}_2^T, \dots, \widetilde{\boldsymbol{F}}_C^T]^T.
	\end{equation}
	Then, the supervector $\widetilde{\boldsymbol{F}}$ 
	can be projected on a low rank subspace using the factor analysis technique to generate the i-vector.
	
	\subsection{NetFV Layer}
	\subsubsection{Fisher Vector}
	Let $\boldsymbol{X}=\{\boldsymbol{x}_i, i=1,2, \dots, L\}$, $\boldsymbol{x}_i \in \mathbb{R}^{D\times1}$, denote the sequence of input features with  $L$ frames, the generation process of the data $\boldsymbol{X}$ is assumed to be modeled by a probability density function $u_\lambda$ with its parameters $\lambda$.
	As argued in~\cite{jaakkola1999exploiting,perronnin2010improving}, the gradient of the log-likelihood describes the contribution of the parameters to the data generation process and can be used as discriminative representation.
	The gradient vector of $\boldsymbol{X}$ w.r.t. the parameters $\lambda$ can be defined as
	\begin{equation}
	\bm{G}^{\boldsymbol{X}}_\lambda = \frac{1}{L}\nabla_\lambda\log{u_\lambda(\boldsymbol{X})}.
	\end{equation}
	A Fisher Kernel is introduced to measure the similarity between two data samples~\cite{jaakkola1999exploiting} and is defined as
	\begin{equation}
	K(\boldsymbol{X}, \boldsymbol{Y}) = (G^{\boldsymbol{X}}_\lambda)'\bm{F}_\lambda^{-1}G^{\boldsymbol{Y}}_\lambda,
	\end{equation}
	where $\boldsymbol{Y}$ is a sequence of features like  $\boldsymbol{X}$, and $\bm{F}_\lambda$ is the Fisher information matrix~\cite{jaakkola1999exploiting} of the probability density function $u_\lambda$:
	\begin{equation}
	\bm{F}_\lambda = E_{\bX \sim u_\lambda}[\nabla_\lambda\log{u_\lambda(\boldsymbol{X})}\nabla_\lambda\log{u_\lambda(\boldsymbol{X})}^T].
	\end{equation}
	Since the $F_\lambda$ is a symmetric and positive semidefinite matrix, we can derive the Cholesky decomposition of the form $\bm{F}_\lambda=\bm{B}_\lambda'\bm{B}_\lambda$, where $\bm{B}_\lambda$ is a lower triangular matrix. In this way, the Fisher kernel $K(\boldsymbol{X}, \boldsymbol{Y})$ is a dot-product between the normalized vectors
	$\bm{\mathcal{G}}^{\bX}_\lambda=\bm{B}_\lambda \bm{G}^{\boldsymbol{X}}_\lambda$,
	where $\bm{\mathcal{G}}^{\bX}_\lambda$ is referred as the Fisher Vector of $\bX$. We choose a $K$-component GMM to model the complex distribution of data, then $u_\lambda(\bx_i) = \sum_{k=1}^{K}\alpha_ku_k(\boldsymbol{x}_i)$, where $\lambda=\{\alpha_k, \boldsymbol{\mu}_k, \boldsymbol{\Sigma}_k, k=1,2,\dots, K\}$ is the parameters set of the GMM. The gradient vector is rewritten as 
	\begin{equation}
	\bm{G}^{\boldsymbol{X}}_\lambda = \frac{1}{L}\sum_{i=1}^{L}{\nabla_\lambda\log{u_\lambda(\bx_i})}.
	\end{equation}
	From the approximation theory of the Fisher information matrix~\cite{jaakkola1999exploiting,perronnin2007fisher,sanchez2013image},  the covariance matrix $\boldsymbol{\Sigma}_k$ can be restricted to a diagonal matrix, that is $\boldsymbol{\Sigma}_k = diag(\boldsymbol{\sigma}_k^2)$, $\boldsymbol{\sigma}_k \in \mathbb{R}^{D\times1}$. 
	Moreover, the normalization of the gradient $G^{\boldsymbol{X}}_\lambda$ by $B_\lambda=F^{-1/2}_\lambda$ is simply a whitening of the dimensions~\cite{perronnin2010improving}. 
	Let $\gamma_i(k)$ be the posterior probability of $\bx_i$ on the $k^\text{th}$ GMM component, 
	\begin{equation}
	\label{eq:post_ocp}
	\gamma_i(k)=\frac{\alpha_k u_k(\bx_i)}{\sum_{c=1}^{K}\alpha_c u_c(\bx_i)}.
	\end{equation}
	The gradient w.r.t. the weight parameters $\alpha_k$ brings little additional information thus it can be omitted~\cite{perronnin2010improving}.
	The remain gradients of $\bx_i$ w.r.t. the mean and standard deviation parameters are derived~\cite{perronnin2010improving} as:
	\begin{equation}
	\label{eq:grad_mean}
	\nabla_{\boldsymbol{\mu}_k}\log{u_\lambda(\boldsymbol{x}_i)} = \frac{1}{\sqrt{\alpha_k}}\gamma_i(k)(\frac{\bx_i-\boldsymbol{\mu}_k}{\bsigma_k}),
	\end{equation} 
	\begin{equation}
	\label{eq:grad_cova}
	\nabla_{\boldsymbol{\sigma}_k}\log{u_\lambda(\boldsymbol{x}_i)} = \frac{1}{\sqrt{2\alpha_k}}\gamma_i(k)(\frac{(\bx_i-\boldsymbol{\mu}_k)^2}{\bsigma_k^2}-1).
	\end{equation} 
	By concatenating the gradients in Eq.~\ref{eq:grad_mean}-\ref{eq:grad_cova}, we get the gradient vector $\bm{\mathcal{G}}_k(\bx_{i})$ with dimension ${2D\times1}$. 
	With the $K$-component GMM, the Fisher Vector of $\bx_i$ is in the form of 
	\begin{equation}
	\bm{\mathcal{G}}(\bx_{i}) = [\bm{\mathcal{G}}_1(\bx_{i})^T, \bm{\mathcal{G}}_2(\bx_{i})^T, \dots, \bm{\mathcal{G}}_K(\bx_{i})^T]^T,
	\end{equation}
	which is a high dimensional vector in $\mathbb{R}^{2DK\times1}$.
	Finally, for the utterance-level representation, the Fisher Vector of the whole sequence $\boldsymbol{X}$ is approximated by a  mean pooling on all $\bm{\mathcal{G}}(\bx_{i})$, 
	\begin{equation}
	\label{eq:fv}
	\bm{\mathcal{G}}^{\bX}_\lambda= \frac{1}{L}\sum_{i=1}^{L}\bm{\mathcal{G}}(\bx_{i}).
	\end{equation}
	
	\subsubsection{NetFV}
	Once the FV codebook is trained, the parameters $\lambda$ of the traditional FV are fixed and can't be jointly learnt with other modules in the end-to-end system. 
	To address this issue, as proposed in ~\cite{tang2016deep}, two simplifications are made to original FV: 1) Assume all GMM components have equal weights. 2) Simplify the Gaussian density $u_k(\boldsymbol{x}_i)$ to
	\begin{equation}
	\setlength{\medmuskip}{0mu}
	u_k(\boldsymbol{x}_i) = \frac{1}{\sqrt{(2\pi)^{D}} }\exp\{-\frac{1}{2}(\boldsymbol{x}_i-\boldsymbol{\mu}_k)^T\boldsymbol{\Sigma}_k^{-1}(\boldsymbol{x}_i-\boldsymbol{\mu}_k)\}.
	\end{equation}
	Let $\boldsymbol{w}_k=1/\bsigma_k$ and $\boldsymbol{b}_k = -\bmu_k$, and with the assumption of $\boldsymbol{\Sigma}_k = diag(\boldsymbol{\sigma}_k^2)$, the gradients in Eq.~\ref{eq:grad_mean}, Eq.~\ref{eq:grad_cova} and the posterior probability  $\gamma_i(k)$ in Eq.~\ref{eq:post_ocp} are respectively rewritten as the final form of NetFV~\cite{tang2016deep}:
	\begin{equation}
	\label{eq:grad_mean_final}
	\nabla_{\boldsymbol{\mu}_k}\log{u_\lambda(\boldsymbol{x}_i)} = \gamma_i(k)[\boldsymbol{w}_k\odot(\bx_i+\boldsymbol{b}_k)],
	\end{equation} 
	\begin{equation}
	\label{eq:grad_cova_final}
	\nabla_{\boldsymbol{\sigma}_k}\log{u_\lambda(\boldsymbol{x}_i)} = \frac{1}{\sqrt{2}}\gamma_i(k)[(\boldsymbol{w}_k\odot(\bx_i+\boldsymbol{b}_k))^2-1],
	\end{equation} 
	\begin{equation}
	\setlength{\medmuskip}{1mu}
	\setlength{\thickmuskip}{0mu}
	\label{eq:post_ocp_final}
	\gamma_i(k) = \frac{\exp\{-\frac{1}{2}(\boldsymbol{w}_k\odot(\bx_i+\boldsymbol{b}_k))^T(\boldsymbol{w}_k\odot(\bx_i+\boldsymbol{b}_k))\}}{\sum_{c=1}^{K}\exp\{-\frac{1}{2}(\boldsymbol{w}_c\odot(\bx_i+\boldsymbol{b}_c))^T(\boldsymbol{w}_c\odot(\bx_i+\boldsymbol{b}_c))\}}.
	\end{equation}
	The three modified equations above are differentiable so the parameters set, i.e., $\{\boldsymbol{w}_k, \boldsymbol{b}_k\}$, can be learnt via the back-propagation algorithm.
	
	\begin{figure}[t]
		\centering
		\includegraphics[width=0.9\linewidth]{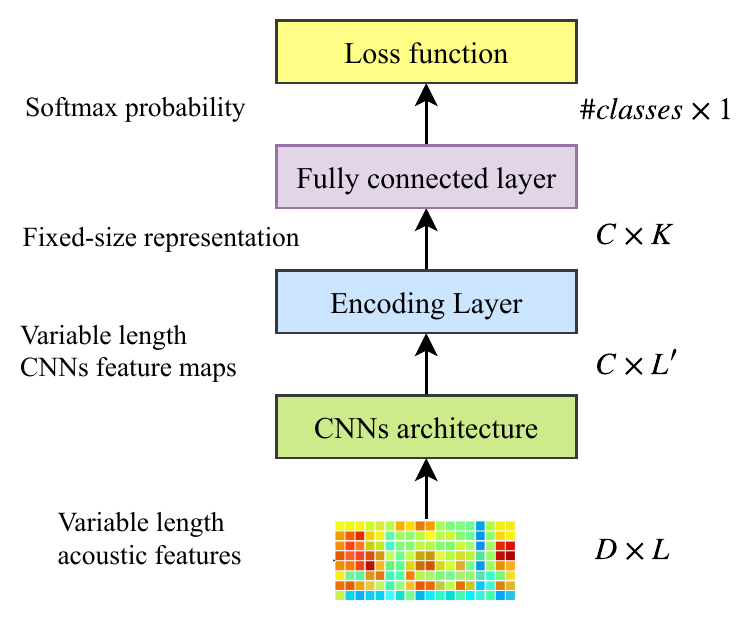}
		\caption{Schematic diagram of end-to-end LRE framework.}
		\label{fig:e2e-lre}
	\end{figure}
	
	\subsection{NetVLAD Layer}
	VLAD is another strategy used to aggregate a set of feature descriptors into a fixed-size representation~\cite{jegou2010aggregating}. 
	With the same inputs $\boldsymbol{X}=\{\bx_i, i=1,2,\dots,L\}$ as FV and $K$ clusters assumed in VLAD, i.e., $\{\bmu_k, k=1,2,\dots, K\}$, the conventional VLAD aligns each $\bx_i$ to a cluster $\bmu_k$. The VLAD fixed-size representation $V \in \mathbb{R}^{K\times D}$ is defined as
	\begin{equation}
	\label{eq:vlad}
	\bm{V}(k) = \sum_{i=1}^{L}\beta_k(\bx_i)(\bx_i-\bmu_k),
	\end{equation}
	where $\beta_k(\bx_i)$ indicates $1$ if $\bmu_k$ is the closest cluster to $\bx_i$ and $0$ otherwise. This discontinuity prevents it to be differentiable in the end-to-end learning pipeline.
	To make the VLAD differentiable, The authors in~\cite{arandjelovic2016netvlad} proposed the soft assignment to  function $\beta_k(\bx_i)$, that is
	\begin{equation}
	\label{eq:beta}
	\beta_k(\bx_i) = \frac{\exp(\boldsymbol{w}_k^T \bx_i + b_k)}{\sum_{c=1}^{K}\exp(\boldsymbol{w}_c^T \bx_i + b_c))},
	\end{equation}
	where $\boldsymbol{w}_k \in \mathbb{R}^{D\times 1}$ and $b_k$ is a scale. By integrating the soft alignment $\beta_k(\bx_i)$ into Eq.~\ref{eq:vlad}, the final form of differentiable VLAD method is derived as
	\begin{equation}
	\label{eq:netvlad}
	\bm{V}(k) = \sum_{i=1}^{L}\frac{\exp(\boldsymbol{w}_k^T \bx_i + b_k)}{\sum_{c=1}^{K}\exp(\boldsymbol{w}_c^T \bx_i + b_c)}(\bx_i-\bmu_k),
	\end{equation}
	which is so-called NetVLAD~\cite{arandjelovic2016netvlad} with the parameters set of $\{\bmu_k, \boldsymbol{w}_k, b_k, k=1,2,\dots,K\}$. 
	The fixed-size matrix in Eq.~\ref{eq:netvlad} is normalized to generate the final utterance-level representations.

	\subsection{Insights into NetFV and NetVLAD}
	Focusing on the GMM supervector (Eq.~\ref{eq:supervector}), the gradient components w.r.t. mean in Fisher Vector (Eq.~\ref{eq:grad_mean}) and the VLAD expression (Eq.~\ref{eq:vlad}), we can found that these three methods calculate the zero-order and first-order statistics to construct fixed dimensional representations in a similar way. The residual vector measures the differences between the input feature and its corresponding component in GMM or cluster in codebooks.
	And all the three aforementioned methods store the weighted sum of residuals. However, they might have different formulas to compute the zero-order statistics.
	As for Fisher Vector, it captures the additional gradient components w.r.t. covariance which can be considered as the second-order statistics.
	Above all, these three encoding methods have theoretical explanations from different perspectives but result in some similar mathematics formulas.
	Motivated by the great success of GMM Supervector, the NetFV and NetVLAD layers theoretically might have good potential in paralinguistic speech attribute recognition tasks.
	
	Compared to the temporal average pooling (TAP) layer, NetFV and NetVLAD layers are capable to heuristically learn more discriminative feature representations in an end-to-end manner while TAP layer may de-emphasize some important information by simple average pooling.
	If the number of clusters $C$ in NetFV or NetVLAD layer is 1, and its mean is zero, the encoding layer is just simplified to TAP layer.

	Our end-to-end framework is illustrated in Fig.~\ref{fig:e2e-lre}. It comprises a CNNs architecture with $C$ output channels, an encoding layer with cluster size $K$ and a fully connected layer. 
	Taken the variable-length features $\boldsymbol{X} \in \mathbb{R}^{D\times L}$ as input, the CNNs structure spatially produces a variable-size feature maps in $\mathbb{R}^{C \times L'}$, where $L'$ is dependent on the input length $L$. 
	The encoding layer then aggregates the feature maps into a fixed-size representation $V$ in $\mathbb{R}^{C \times K}$. And the fully connected layer acts as a back-end classifier.
	All the parameters in this framework are learnt via the back-propagation algorithm.
	

	\section{Experiments}
	\label{sec:experiments}
	\subsection{2007 NIST LRE Closet-set Task}
	The 2007 NIST Language Recognition Evaluation(LRE) is a closed-set language detection task. The training set consists the datasets of Callfriend, LRE 03, LRE 05, SRE 08 and the development part of LRE 07. We split the utterances in  training set into segments with duration in 3 to 120 seconds. This yields about 39000 utterances. 
	In the test set, there are 14 target languages with
	7530 utterances in total. The nominal durations of the testing data are 3s, 10s and 30s.
	
	\subsection{Experimental Setup}
	Raw audio is converted to 7-1-3-7 based 56 dimensional shifted delta coefficients (SDC) feature, and a frame-level energy-based voice activity detection (VAD) selects features corresponding to speech frames. 
	We train a 2048-component GMM-UBM with full covariance and extract 600-dimensional i-vectors followed by the whitening and length normalization. Finally, we adopt multi-class logistic regression to predict the language labels.
	
	For the end-to-end LID systems, the 64-dimension mel-filterbank coefficients feature is extracted along with sliding mean normalization over a window of 300 frames. Afterwards, the acoustic features are fed to a random initialized ResNet-34 networks with 128 output channels to produce the utterance-dependent feature maps. 
	The temporal average pooling (TAP) is adopted as the encoding layer to build the baseline end-to-end system. Meanwhile, the cluster size $K$ ranges from 16 to 128 by step power of $2$ to find out the best parameter setup in NetFV and NetVLAD layers, respectively. 
	The softmax and cross entropy loss are integrated behind the fully connected layer. Finally, a stochastic gradient descent (SGD) optimizer with the momentum 0.9, the weight decay \num{1e-4} and the initial learning rate 0.1 is used for the back-propagation. The learning rate is divided by 10 at the $60^{th}$ and $80^{th}$ epoch.
	
	To efficiently train the system, we set the mini-batch size of 128 in data parallelism over 4 GPUs. For each mini-batch data, a truncated-length $L$ is randomly sampled from $[200, 1024]$, and is used to truncate a segment of continuous $L$-frame features from a $T$-frame utterance. The beginning index of truncation randomly lies in  the interval of $[0, T-L-1]$. Consequently, a mini-batch data with unified-length samples in $\mathbb{R}^{128 \times 64 \times L}$ is loaded and $L$ may change for different mini-batches. 
	In the test phase, all the speeches in 3, 10 and 30 seconds durations are tested one-by-one on the same trained model. No truncation is used for the arbitrary-duration utterances.
	
	\begin{figure}[t]
		\centering
		\includegraphics[width=0.95\linewidth]{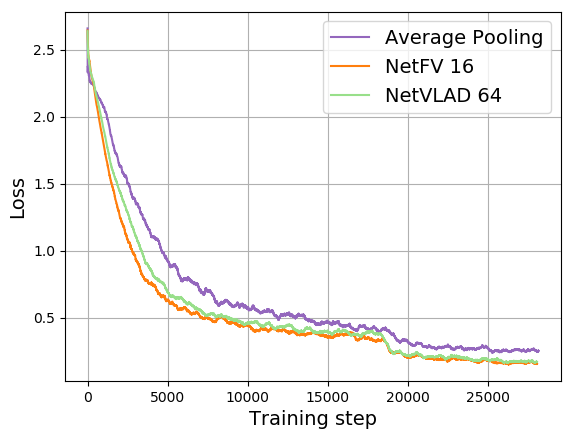}
		\caption{The training loss curves of end-to-end systems.}
		\label{fig:loss-curves}
	\end{figure}

	\subsection{Evaluation}
	The training losses of the end-to-end systems are sliding smoothed with window size of 400, and illustrated in Fig.~\ref{fig:loss-curves}.
	The NetFV and NetVLAD based systems converge faster and reach lower losses than that of the temporal average pooling (TAP) layer.
	With a closer look, NetFV is slightly better than NetVLAD but both of them are competitive in the training phase.
	The language identification results are presented in Table~\ref{tab:result_lre07}. The performance is measured in the metrics of the average detection cost $C_{avg}$ and equal error rate (ERR). 
	From the Table~\ref{tab:result_lre07}, the end-to-end systems including TAP, LDE, NetFV and NetVLAD layers significantly outperform the conventional GMM i-vector baseline. It shows that the proposed end-to-end framework for the LID task is feasible and effective.
	The results of the systems based on the TAP and the remarkable LDE layers are provided in our early works~\cite{caiwch2018lde}.
	
	Moreover, we step further to compare the performances of the TAP, NetFV and NetVLAD encoding layers. 
	Both NetFV and NetVLAD based systems achieve much lower $C_{avg}$ and EER than the ones with TAP layer. Especially on the long utterances (30s), the best $C_{avg}$ and EER of NetVLAD could be relatively reduced by $27.87\%$ and $71.98\%$ respectively w.r.t. the results of TAP.
	If we concentrate on the NetFV and NetFV only, the performances are generally getting better while the cluster size ranges 16 to 64, and start to degrade when the cluster size reaches 128.
	Therefore, larger cluster size in the encoding layer may enhance the capacity of networks, however, more data and training epochs may be required as well.
	In addition, the TAP based system shows the accuracy rates of 75.49\%, 89.71\% and 93.56\% on the 3s, 10s and 30s test set respectively while the NetVLAD based system improves the accuracies to 76.14\%, 91.43\% and 96.85\%.
	Overall, NetVLAD is slightly superior to NetFV in the test phase and achieves the best performance when the cluster size is 64. 
	
	Despite the best result corresponding to the NetVLAD is slightly inferior to that of the LDE layer, the performances of the NetFV, NetVLAD and LDE layers are comparable. What's more, these three powerful encoding methods are complementary. With the cluster size is 64, we fuse the three systems based on the NetFV, NetVLAD and LDE respectively at the score level. And as shown in the Table~\ref{tab:result_lre07}, the score level fusion system further reduces the $C_{avg}$ and EER significantly.
	\begin{table}[t]
		\caption{Performances on 2007 NIST LRE task}
		\label{tab:result_lre07}
		\centering
		\setlength{\tabcolsep}{3pt}
		\begin{tabular}{ l c|c|c}
			\toprule
			\multirow{2}*{\textbf{System description}} & 
			\multicolumn{3}{c}{$C_{avg}(\%)/EER(\%)$}  \\\cline{2-4}
			& 3s & 10s & 30s \\
			\midrule
			GMM i-vector & 20.46/17.71 & 8.29/7.00 & 3.02/2.27\\
			\hline
			ResNet34 TAP	&  9.24/10.91 & 3.39/5.58 &  1.83/3.64 \\
			\hline
			ResNet34 LDE 64 &   \textbf{8.25}/\textbf{7.75}  &  \textbf{2.61}/\textbf{2.31} &  \textbf{1.13}/\textbf{0.96} \\
			\hline
			ResNet34 NetFV 16    & 9.47/9.04 & 2.96/2.59   & 1.31/1.08 \\  
			ResNet34 NetFV 32   & 8.95/8.37 & 2.88/2.49   & 1.35/1.31 \\
			ResNet34 NetFV 64   & 8.91/8.26 & 2.88/2.74   &  1.19/1.15 \\ 	      
			ResNet34 NetFV   128   & 9.05/8.64 & 2.91/2.72   & 1.27/1.34 \\
			\hline
			ResNet34 NetVLAD 16 &  \textbf{8.23}/\textbf{8.06}  &  2.90/2.62 &  1.36/1.17 \\
			ResNet34 NetVLAD 32 &  8.87/8.58  &  3.10/2.50 &  1.46/1.15 \\
			ResNet34 NetVLAD 64 &  8.59/8.08  &  \textbf{2.80}/\textbf{2.50} &  \textbf{1.32}/\textbf{1.02} \\
			ResNet34 NetVLAD 128 &  8.72/8.44  &  3.15/2.76 &  1.53/1.14 \\
			\hline
			Fusion system  & \textbf{6.14}/\textbf{6.86} & \textbf{1.81}/\textbf{2.00} & \textbf{0.89}/\textbf{0.92}\\
			\bottomrule
		\end{tabular}
	\end{table}

\section{Conclusions}
	\label{sec:conclusions}
	
	In this paper, we apply these two encoding methods in our end-to-end LID framework to investigate the feasibility and performance. 
	The NetFV and NetVLAD layers are more powerful encoding techniques with learnable parameters and are able to encoding the variable-length sequence of features into a fixed-size representation. We integrate them to  a flexible end-to-end framework for the LID task and conduct experiments on the NIST LRE07 task to evaluate the methods. Promising experimental results show effectiveness and great potential of NetFV and NetVLAD in the LID task. This end-to-end framework might also work for other paralinguistic speech attribute recognition tasks, which will be our further works.

	\bibliographystyle{IEEEtran}
	
	\bibliography{mybib}

\end{document}